\documentclass[reprint,showpacs,preprintnumbers,amsmath,amssymb,prx,floatfix,superscriptaddress,aps]{revtex4-2}
\usepackage{blindtext}
\usepackage{graphicx}
\usepackage[utf8]{inputenc} 
\usepackage{setspace}
\usepackage{dcolumn}
\usepackage{bm}
\usepackage{color}
\usepackage{SIunits}
\usepackage[normalem]{ulem}
\usepackage{url}
\usepackage{comment}
\usepackage[left, modulo, pagewise]{lineno}

\begin{document}

\title{Fundamental limit of intensity scaling for harmonic generation on plasma mirrors}

\author{Philipp Sikorski}
\thanks{These authors contributed equally to this work.}
\affiliation{Université Paris-Saclay, CEA, LIDYL, 91191 Gif-sur-Yvette, France}

\author{Baptiste Groussin}
\thanks{These authors contributed equally to this work.}
\affiliation{Laboratoire d’Optique Appliquée (LOA), CNRS, École polytechnique, ENSTA, Institut Polytechnique de Paris, Palaiseau, France}

\author{Aodhan McIlvenny}
\affiliation{Lawrence Berkeley National Laboratory, 1 Cyclotron Road, Berkeley, California, 94720, USA}

\author{Kosta Oubrerie}
\author{Pierre Bartoli}
\affiliation{Université Paris-Saclay, CEA, LIDYL, 91191 Gif-sur-Yvette, France}

\author{Lieselotte Obst-Huebl}
\affiliation{Lawrence Berkeley National Laboratory, 1 Cyclotron Road, Berkeley, California, 94720, USA}

\author{Anthony Vazquez}
\author{Lulu Russell}
\affiliation{Lawrence Berkeley National Laboratory, 1 Cyclotron Road, Berkeley, California, 94720, USA}
\affiliation{University of California, Berkeley, California, 94720, USA }

\author{Tirtha Mandal}
\thanks{Current address: Raja Ramanna Centre for Advanced Technology, Indore 452013, India}
\affiliation{Lawrence Berkeley National Laboratory, 1 Cyclotron Road, Berkeley, California, 94720, USA}

\author{Kei Nakamura}
\author{Anthony J. Gonsalves}
\author{Cameron G. R. Geddes}
\affiliation{Lawrence Berkeley National Laboratory, 1 Cyclotron Road, Berkeley, California, 94720, USA}

\author{Luca Fedeli}
\affiliation{Université Paris-Saclay, CEA, LIDYL, 91191 Gif-sur-Yvette, France}

\author{Henri Vincenti}
\email{henri.vincenti@cea.fr}
\affiliation{Université Paris-Saclay, CEA, LIDYL, 91191 Gif-sur-Yvette, France}

\author{Adrien Leblanc}
\email{adrien.leblanc@ensta-paris.fr}
\affiliation{Laboratoire d’Optique Appliquée (LOA), CNRS, École polytechnique, ENSTA, Institut Polytechnique de Paris, Palaiseau, France}

\date{\today}

\begin{abstract}

Plasma mirrors have the potential to considerably boost the intensity of ultra-intense laser pulses through high-order harmonic generation, offering a promising route to the extreme fields required for probing the strong-field regime of quantum electrodynamics (SF-QED). This prospect relies on the favorable scaling of harmonic generation with increasing laser intensity, a longstanding assumption that has remained untested at extreme intensities. Combining the first measurements of Doppler harmonic generation approaching \(10^{22}\,\mathrm{W/cm^2}\) with advanced particle-in-cell modeling, our study reveals fundamental limitations to this scaling. We notably observe a strong reduction of the harmonic yield as the laser intensity is increased, and show using simulations that this originates from the interaction of the sub-picosecond laser pedestal with the plasma mirror. This pedestal drives a rapid evolution of the plasma surface, through radiation-pressure compression and surface modulations, which progressively suppress harmonic generation. By extending our simulations up to intensities of \(10^{23}\,\mathrm{W/cm^2}\), we show that this evolution accelerates strongly with increasing intensity, eventually leading to plasma mirror disruption on the timescale of the main laser pulse itself, making pulse duration, in addition to temporal contrast, a second key parameter for efficient harmonic generation. These results reveal the fundamental limits of the intensity scaling of harmonic generation from plasma mirrors and establish the conditions required to preserve efficient harmonic generation at extreme intensities, providing critical guidance for their use as intensity boosters toward the SF-QED regime.
\end{abstract}

\pacs{Valid PACS appear here}

\maketitle

\section{Introduction}
The advent of chirped-pulse amplification \cite{StricklandOptComm1985} has enabled light intensities high enough to drive matter into the relativistic regime, paving the way for the fields of relativistic optics \cite{MourouRevModPhys2006} and ultra-high intensity science \cite{EsareyRevModPhys2009, TeubnerRevModPhys2009, DiPiazzaRevModPhys2012, MacchiRevModPhys2013}. In this context, plasma mirrors \cite{KapteynOptLett1991, DoumyPRE2004} have emerged as key optical elements for manipulating ultra-intense light \cite{ThauryNatPhys2007, DromeyPRL2007, DromeyNatPhys2009, Nakatsutsumi:10, vincenti2014optical,denoeud2017interaction, LeblancNatPhys2017}. They are formed when a tightly focused femtosecond (fs) laser pulse ionizes a solid target, producing a dense plasma that reflects the incident light. When the plasma–vacuum interface has a density gradient of scale length $L$ much shorter than the laser wavelength $\lambda$, the plasma surface behaves as a mirror of excellent optical quality \cite{ScottNJP2015, KahalyPRL2013, chopineau2019identification}.

Plasma mirrors have proven essential for applications ranging from laser-driven ion acceleration \cite{NeelyAppPhysLett2006, CeccottiPRL2007, MacchiRevModPhys2013} to high-order harmonic generation (HHG) \cite{TeubnerRevModPhys2009, GaoChinOptLett2017}, the latter being central to the present work. Harmonics up to the extreme ultraviolet (XUV) spectral region can be generated via the Doppler effect by the relativistic oscillations of the surface of a plasma mirror irradiated at ultra-high intensities \cite{BulanovPoP1994, LichtersPoP1996, TeubnerRevModPhys2009, ThauryJPB2010, YunRevModPlasmaPhys2025}. Initially envisioned as a source to perform attosecond pump-probe experiments \cite{NomuraNatPhys2009} and probe relativistic laser–plasma interactions \cite{KrushelnickPPCF2002}, Doppler harmonic generation has recently attracted renewed interest as a promising route to boost PW-class laser intensities by several orders of magnitude \cite{GordienkoPRL2005, gonoskov2011ultrarelativistic, VincentiPRL2019, quere2021reflecting, ChopineauNatPhys2021, vincenti2023plasma}. Simulations predict that successfully operating plasma mirrors in the multi PW-regime could yield boosted laser intensities well above $10^{25} ~\textrm{W/cm}^2$, opening a pathway \cite{KarbsteinPRL2019, FedeliPRL2021, SainteMarieNJP2022, HuiAIP2023, ZaimPRL2024} to probe the strong-field regime of quantum electrodynamics (QED)\cite{DiPiazzaRevModPhys2012, GonoskovRevModPhys2022}. This intensification arises from a relativistic spatiotemporal compression: as the laser drives the plasma mirror surface to oscillate at relativistic velocities, each optical cycle experiences temporal compression and Doppler upshifting into the XUV, while the plasma curvature --- induced by the laser radiation pressure --- focuses the emitted radiation.

The use of plasma mirrors in such applications hinges on preserving a sharp gradient scale length. This is a real challenge as it requires a high laser contrast, commonly defined as the ratio of the peak intensity to that of the nanosecond-scale pedestal preceding the main pulse. This pedestal can pre-ionize and heat the target, driving unwanted plasma expansion and degrading the interface, unless properly suppressed \cite{ThauryNatPhys2007, chopineau2019identification}. To this end, various contrast-enhancement techniques have been developed, the most common involving ``optical switches'', also known as plasma mirrors, that transmit the pedestal but become ionized at a threshold intensity, acting as ultrafast shutters \cite{KapteynOptLett1991, DromeyRevSciInst2004, DoumyPRE2004, NomuraNJP2007, HorleinNJP2008}. By achieving laser contrast better than $10^{12}$ \cite{Czapla2025, ObstPPCF2018,ChoiOptLett2020,AkiraHPLSE2022}, these devices have proven highly effective for driving relativistic plasma mirrors in applications using high-power lasers \cite{DansonHPLSE2019} at intensities well below $10^{21} ~ \textrm{W/cm}^2$, with only two works \cite{DollarPRL2013,TimmisNature2026} slightly above $10^{21} ~ \textrm{W/cm}^2$. 

\begin{figure}[t]
    \includegraphics[width=\columnwidth]{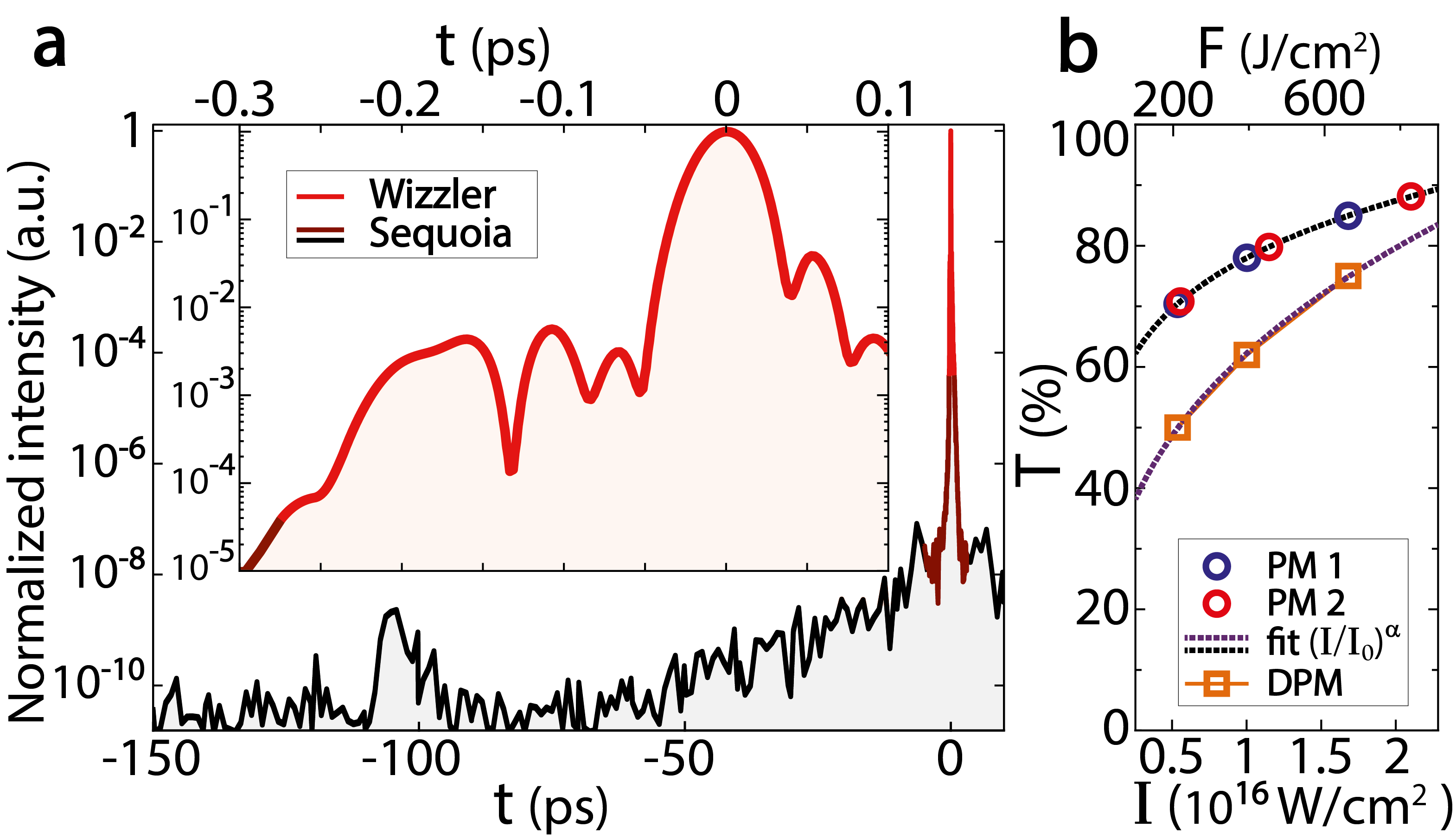}
    \caption{(a) Laser temporal profile measured with a Sequoia cross-correlator: -150ps window / 0.9ps resolution in black, $\pm 5$ps window / 0.1ps resolution in red. Inset: sub-300fs Wizzler measurement before the DPM. (b) DPM transmission as a function of the maximum intensity and fluence incident on the first PM surface (orange markers). The deduced transmissions of each plasma mirror (red and blue circles) are shown as a function of the intensity on each PM surface (black and purple lines), obtained from a power-law fit (see Sup. Mat. \cite{SuppMat})}
    \label{fig:laser_contrast_DPM_transmission}
\end{figure}

Here, we report the first study of Doppler harmonic generation from plasma mirrors at intensities substantially exceeding $10^{21} ~\textrm{W/cm}^2$ and approaching $10^{22} ~\textrm{W/cm}^2$, using a PW-class laser. By systematically measuring the XUV harmonic properties as a function of laser and plasma mirror parameters, we show that for peak intensities above a few $10^{21} ~\textrm{W/cm}^2$ the laser pedestal at the sub-picosecond (ps) time scale becomes a critical factor governing plasma mirror quality and harmonic generation efficiency. Building on these experimental observations, we conducted a detailed numerical study to elucidate how the plasma mirror dynamics under intense laser irradiation affect the HHG efficiency. Crucially, our simulations reveal that, at sufficiently high intensities, the duration of the main laser pulse itself becomes a limiting factor. Together, these results provide the first comprehensive picture of the laser contrast and pulse-duration requirements for efficient HHG with PW and multi-PW class lasers, establishing key guidelines for the success of applications employing plasma mirrors at ultra-high intensities.

\begin{figure}[tb]
    \includegraphics[width=\columnwidth, height=0.6\textheight,keepaspectratio]{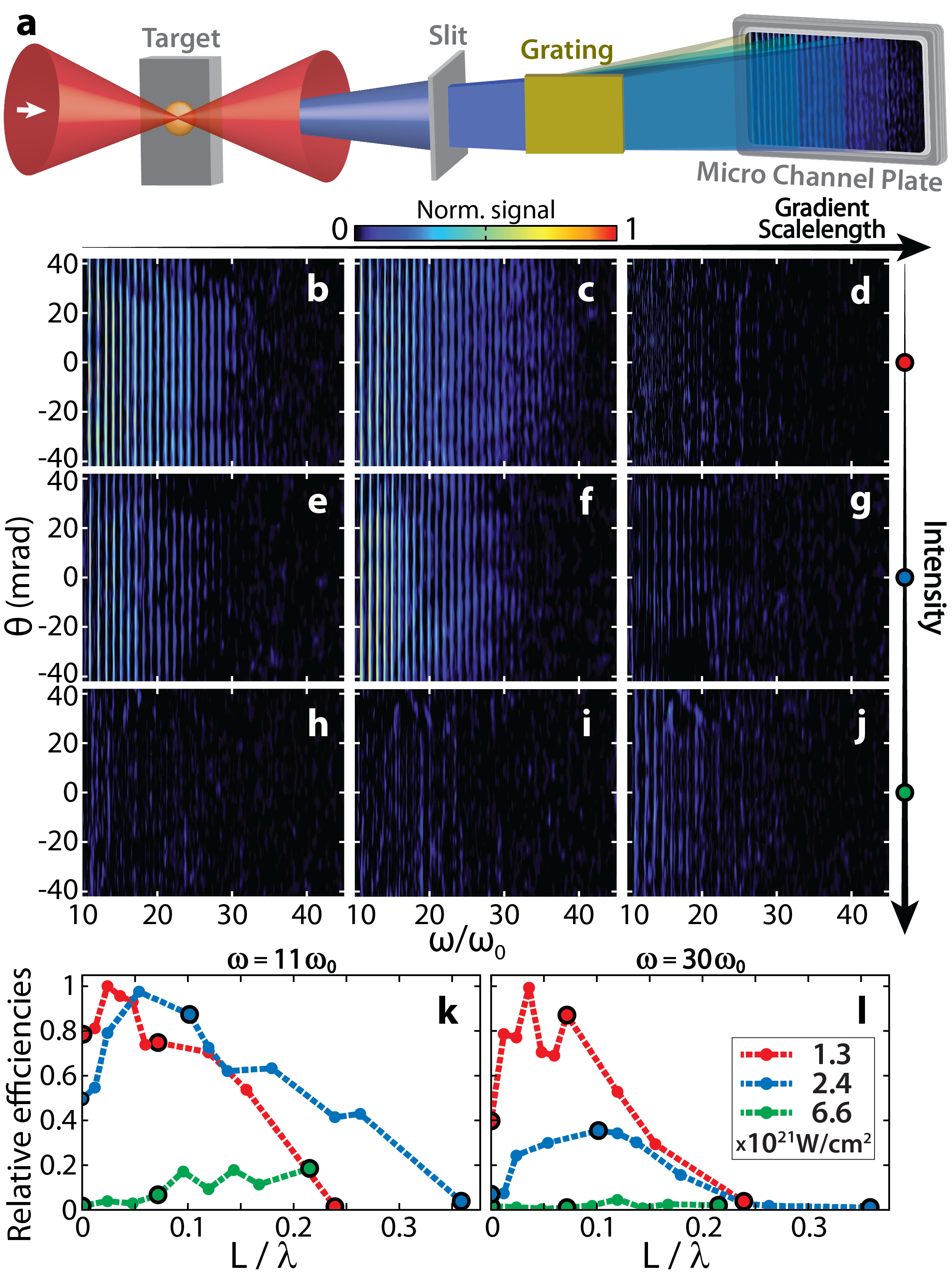}
    \caption{(a) Experimental scheme: the p-polarized laser pulse is focused onto the solid target and the XUV spectrum is measured using a $2$-mm entrance slit,  corresponding to an angular acceptance of 6 mrad, and a curved XUV grating imaging the HHG source (in the dispersive direction) onto a micro-channel-plate detector coupled to a phosphor screen, the emission of which is recorded by a camera.
    (b-j) Angularly resolved XUV spectra for different on-target laser intensities: $1.3 \times 10^{21} ~ \textrm{W/cm}^2$ (b–d), $2.4 \times 10^{21}~ \textrm{W/cm}^2$ (e–g), and $6.6 \times 10^{21} ~ \textrm{W/cm}^2$ (h–j), and for different gradient scale lengths: short ($L=0$, first column), intermediate ($L\approx \lambda/11$, second column), and long ($L > \lambda/5$, third column). All spectra are normalized to the maximum value across the entire dataset. (k-l) Relative efficiency of the $\textrm{11}^{th}$ (k) and $\textrm{30}^{th}$ (l) harmonics as a function of the plasma gradient length, calculated as the sum of the signal within a 6mrad x 85mrad angular window. For each harmonic order, the efficiencies are normalized to the maximum value across the entire dataset. Data points corresponding to panels (b–j) are marked with a black contour.}
    \label{fig:exp_results}
\end{figure}

\section{HHG experiments at intensities approaching $10^{22} ~\textrm{W/cm}^2$} \label{sec:exp}

\subsection {Experimental set-up and harmonic measurements}
The experiment was conducted at the BELLA PW laser facility \cite{nakamura2017diagnostics} at Lawrence Berkeley National Laboratory (LBNL). Pulses with an energy up to 31~J after compression down to $37$-fs Full Width Half Maximum (FWHM) duration were used. The pulse temporal profile is presented in Fig. \ref{fig:laser_contrast_DPM_transmission}(a). Temporal contrast enhancement was achieved using a double-plasma-mirror (DPM) system (see design in the Supplementary Material (Sup. Mat.) \cite{SuppMat}). The total DPM transmission, along with the inferred transmission of each individual plasma mirror (see Sup. Mat. \cite{SuppMat}) and the corresponding fluences/intensities on their surfaces, is presented in Fig. \ref{fig:laser_contrast_DPM_transmission}(b). These measurements show excellent agreement with previous reports \cite{DoumyPRE2004, InoueApplOpt2016, ChoiOptLett2020, ZieglerSciRep2021, NishiuchiHPLSE2022} (see Sup. Mat. \cite{SuppMat}).

After the DPM, the beam is focused down to a $2.8\,$\micro m FWHM focal spot onto a SiO$_2$ polished target with an angle of incidence of $\approx 45^\circ$, using a $0.5\ $m off-axis parabola in the BELLA iP2 target chamber. A $9$ mm pick-off mirror positioned upstream of the final parabola samples a small portion of the beam, which is loosely focused to a $48\,$\micro m FWHM spot at $2$–$6 \times 10^{15}~\textrm{W/cm}^2$
 with an adjustable delay relative to the main pulse for pre-ionizing the target and controlling the gradient scalelength during the interaction to optimize HHG yield \cite{KahalyPRL2013}. Previous measurements performed at comparable laser fluences reported plasma expansion velocities ranging from $15$ to $30$ nm/ps \cite{bocoum2015spatial}, consistent with earlier studies \cite{KahalyPRL2013,chopineau2019identification}.

As illustrated in Fig. \ref{fig:exp_results}(a), an XUV spectrometer was placed along the specular direction of the reflected beam to record the angular profile of harmonics above the $11^{\textrm{th}}$ order with an angular acceptance of $6$ mrad horizontally (defined by the entrance slit) and $85$ mrad vertically (limited by the grating width).  The spectrometer thus samples the XUV emission within a limited angular window around the specular direction.

\begin{figure*}[t]
    \includegraphics[width=\textwidth]{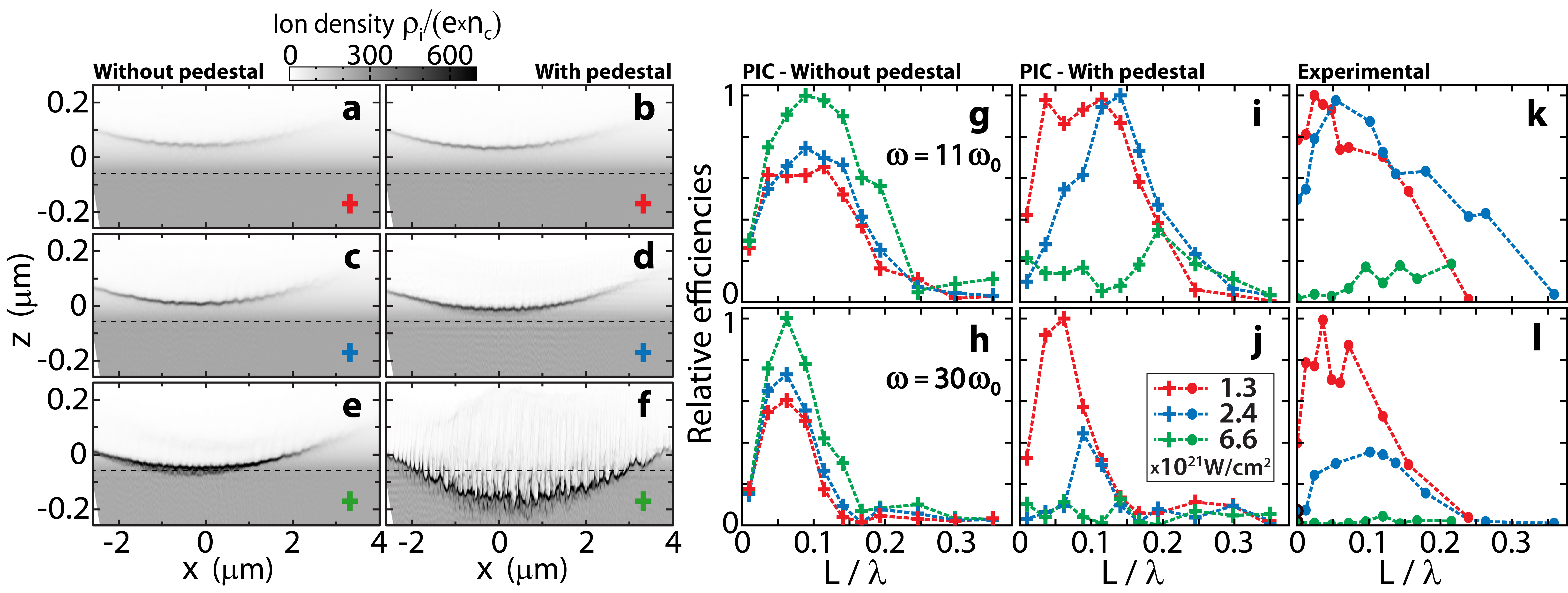}
    \caption{High-resolution PIC simulations corresponding to the experimental cases. (a–f) For an initial plasma gradient length of $\lambda/10$: target ion density (in units of \(\rho_i/(e\,n_c)\), where \(\rho_i\) is the ion charge density and \(n_c=1.7\times10^{21}\,\mathrm{cm}^{-3}\) is the critical electron density) at the peak of the laser pulse for intensities of $1.3 \times 10^{21} ~ \textrm{W/cm}^2$ (a–b), $2.4 \times 10^{21} ~ \textrm{W/cm}^2$ (c–d), and $6.6 \times 10^{21} ~ \textrm{W/cm}^2$ (e–f). First column (a,c,e): idealized 37 fs FWHM Gaussian laser pulse (no pedestal); second column (b,d,f): realistic laser temporal profile (with pedestal). In each panel, the black dashed line delimits the initial boundary between the plasma gradient and the bulk region. (g-l) Harmonic generation efficiency for the $\textrm{11}^{th}$ (first row) and $\textrm{30}^{th}$ (second row) orders as a function of plasma scale length, calculated as the sum of the harmonic signal collected by the detector (or within the same 85 mrad angular window for numerical data), for the three intensities on target: 1.3 (red lines), 2.4 (blue), and 6.6 $\times 10^{21} ~ \textrm{W/cm}^2$ (green). Harmonic efficiency from PIC simulations without pedestal, first column (g-h), with realistic laser temporal profile, second column (i-j), compared with the experimental harmonic efficiencies, last column (k-l). For each panel, the efficiencies are normalized with respect to the maximum value. }
    \label{fig:exp_vs_num_res}
\end{figure*}

XUV spectra were recorded for approximately ten gradient scale lengths between 0 and 0.4$\lambda$ and for three laser energies: $10$ J, $18$ J, and $31$ J on DPM, corresponding to $5$ J, $11$ J, and $23$ J on target and intensities of $1.3$, $2$, and $7\times 10^{21} ~\textrm{W/cm}^2$, respectively. For each intensity, three representative angularly resolved spectra are shown in Figs. \ref{fig:exp_results} (b–j), corresponding to short ($L=0$), intermediate ($L\approx \lambda/11$), and long ($L> \lambda/5$) gradients. The XUV signals integrated over the spectrometer's angular window are plotted as a function of the gradient scale length for harmonic orders $\textrm{11}^{th}$ and $\textrm{30}^{th}$ in Figs. \ref{fig:exp_results}(k) and (l). For the lower laser intensities, the harmonic emission exhibits the well-established dependence on the plasma density gradient scale length, with the harmonic yield first increasing with gradient length, reaching an optimum around $L\approx 01\lambda$, and then decreasing for longer gradients. In addition, our results reveal a strong decrease of the harmonic emission efficiency with increasing laser intensity: high-order harmonics up to the $\textrm{40}^{th}$ and $\textrm{30}^{th}$ orders are observed at optimized gradients for $1.3$ and $2.4\times 10^{21} ~\textrm{W/cm}^2$, respectively, but a very low harmonic signal is detected at $6.6\times 10^{21} ~\textrm{W/cm}^2$.

\subsection {Origin of the HHG decline at high intensities}
In order to gain insights on how HHG is affected by the interplay of the laser temporal contrast and of the plasma gradient scale length at increased intensities, we carried out an extensive numerical simulation campaign, using the high-performance Particle-In-Cell \cite{BirdsallBook2018, ArberPPCF2015} (PIC) code WarpX v 25.09 \cite{FedeliSC22, WarpXv25.09}. Simulations were carried out in 2D geometry to allow exploring a large parameter space (see Sup. Mat. \cite{SuppMat} for a detailed description of the numerical setup, and also references \cite{VayJCP2013, VincentiCPC2016, BlaclardPRE2017, ThevenetLasy2024} therein). We performed several parameter scans, presenting the results of over 850 simulations in the following sections.

By comparing the experimental results with kinetic plasma simulations, we show that the reduction of XUV emission efficiency with higher laser intensities originates from the sub-ps pedestal preceding the main pulse. As shown in Fig. \ref{fig:laser_contrast_DPM_transmission}(a), the temporal profile exhibits a sub-300-fs pedestal with an average contrast of $2.7\times10^{-3}$ between $-220$ and $-50$ fs (not filtered by the DPM), typical of lasers in this class \cite{KiriyamaOptLett2018,NishiuchiHPLSE2022} and commonly attributed to fine spectral modulations \cite{KiriyamaHPLSE2021}. 

 Plasma mirrors typically ionize at intensities on the order of few $10^{13} ~\textrm{W/cm}^2$ \cite{KapteynOptLett1991, DromeyRevSciInst2004, DoumyPRE2004, ObstPPCF2018}. Assuming a DPM ionization triggering intensity of $2\times 10^{13} ~\textrm{W/cm}^2$, the temporal profiles of the laser reflected from the DPM can be inferred (see Sup. Mat. \cite{SuppMat}). For the three experimental cases with energies after compression of 10, 18, and 31 J, this analysis yields sub-ps contrasts with average values of  $2.8 \times 10^{-5}$, $1.1 \times 10^{-3}$, and $1.8 \times 10^{-3}$ during temporal windows of 110 fs,  175 fs, and 200 fs prior to the pulse peak power, respectively. These contrast levels correspond to target irradiation before the arrival of the main pulse at intensities of  $3.6 \times 10^{16} ~\textrm{W/cm}^2$ over 110 fs, $2.6 \times 10^{18} ~\textrm{W/cm}^2$ over 175 fs, and $1.1 \times 10^{19} ~\textrm{W/cm}^2$ over 200 fs, respectively. 
 
 Reaching relativistic intensities well before the main pulse, this pedestal can significantly degrade the optical quality of the high-intensity plasma-mirror, as shown in Figs. \ref{fig:exp_vs_num_res}(a–f).First, the pedestal can induce a larger denting of the target, and shortens the controlled density gradient formed by the pre-pulse (see Fig. \ref{fig:exp_vs_num_res}(e) vs (f)), shifting the interaction conditions from those required for optimal HHG, which could explain why longer initial scale lengths are required to optimize the HHG efficiency. Moreover, the pedestal can induce a surface density modulation (see Fig. \ref{fig:exp_vs_num_res}(f)), which degrades the optical quality of the plasma mirror. Such modulation is reminiscent of transverse Weibel-type instabilities reported in similar laser-solid interaction scenarios\cite{QuinnPRL2012, Ruyer_laser_induced_shocks_PoP2015, GrassiPRE2017, SchoenwaelderNatComm2026}, whose contribution is further supported by the magnetic filamentation observed in our simulations, as discussed in section  \ref{sec:PM_dynamics_HHG}. This shows that the collapse of HHG efficiency can be attributed to a combination of the aforementioned effects.

\begin{figure*}[t]
    \includegraphics[width=\textwidth]{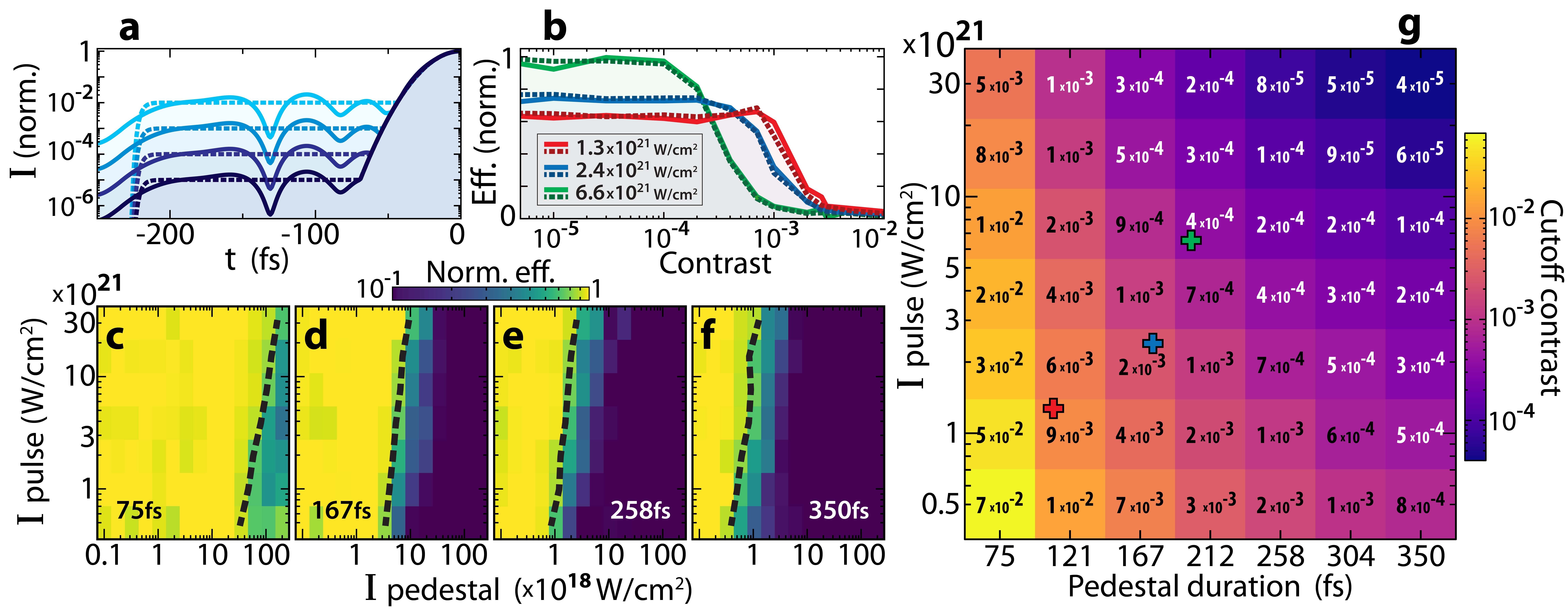}
    \caption{    
    PIC simulations to investigate the impact of sub-ps contrast on HHG efficiency of the $\textrm{11}^{th}$ order for a plasma scale length of $\lambda/10$ (optimizing HHG in the absence of pedestal, see Fig.\ref{fig:exp_vs_num_res}(g)). For a more general study of HHG efficiency, no angular filtering corresponding to a specific detector window is applied here. (a,b) HHG efficiency, panel (b), for the three experimental main pulse peak intensities as a function of the average sub-220 fs temporal contrast, panel (a), using a realistic temporal profile (the one of the laser before the DPM), solid lines, and a simplified 220 fs super-Gaussian profile (order 60), dashed lines. (c–f) HHG efficiency versus sub-350fs pedestal intensity and main pulse intensity for a super-Gaussian pedestal whose duration is 75 fs (c), 167 fs (d), 258 fs (e), and 350 fs (f). Black dashed lines indicate the pedestal intensity cutoff, defined as where the HHG efficiency drops below $70\%$ of the pedestal-free case. (g) Deduced cutoff contrasts for controlled laser–plasma interaction as a function of pedestal duration and on-target main pulse intensity. Colored crosses mark the three probed experimental conditions.   
    }
    \label{fig:sim_constrast_effect}
\end{figure*}

In the absence of a pedestal, simulations show that the harmonic yield increases with the intensity on target, and that the harmonic efficiency is maximized for a plasma density gradient scale length of $L \sim \lambda/10$, independently of the peak laser intensity, see Figs. \ref{fig:exp_vs_num_res}(g–h), which is in agreement with previous numerical studies \cite{VincentiPRL2019}. When the pedestal is included, the simulations indicate that the harmonic yield drops significantly with increasing intensity, leading to a collapse of harmonic generation at $6.6 \times 10^{21} ~\textrm{W/cm}^2$. In addition, the optimal scale length shifts toward larger values as the peak intensity increases (see Figs. \ref{fig:exp_vs_num_res} (i–j)). These results are in very good agreement with the experimental observations in Figs. \ref{fig:exp_vs_num_res} (k-l).

\subsection {Sub-ps contrast requirements for HHG at PW}
In the light of these observations, we now use PIC simulations (see detailed parameters in \cite{SuppMat}) to assess the threshold laser contrast required for efficient HHG as a function of the peak laser intensity on target and the sub-ps pedestal duration, see Fig. \ref{fig:sim_constrast_effect}. For more general conclusions, panels (a–b) show that realistic and simplified super-Gaussian pedestals with the same average contrast have equivalent effects, so the systematic study of pulse intensity, pedestal duration, and contrast is carried out considering a super-Gaussian pedestal.
As the peak intensity increases, a higher temporal contrast is required for a given pedestal duration, Fig. \ref{fig:sim_constrast_effect}(b). Similarly, for longer pedestal durations, the required contrast rises rapidly, Figs. \ref{fig:sim_constrast_effect}(c–f). The laser contrast threshold as a function of both pedestal duration and peak intensity is summarized in Fig. \ref{fig:sim_constrast_effect}(g), providing an important guidelines for optimizing HHG with plasma mirrors at extreme intensities. 

These guidelines notably indicate that DPM systems currently available at PW-class facilities, primarily designed to improve the nanosecond temporal contrast, are not yet suited for a precise control of the sub-ps laser–plasma interaction at intensities beyond $10^{21} ~\textrm{W/cm}^2$. In this regard, our numerical work provides essential guidelines for the tuning of DPM systems to enhance laser contrast at such extreme intensities. 

A recent experiment \cite{TimmisNature2026} demonstrated that, already at intensities of $8\times 10^{20},\mathrm{W/cm^2}$, the sub-picosecond contrast can be substantially improved by replacing the first anti-reflection-coated  plasma mirror of the DPM with an uncoated substrate. This increases its switching threshold, allowing the rising pedestal to be transmitted up to higher intensities and resulting in substantially enhanced harmonic conversion efficiency. Here, we introduce a complementary approach that provides continuous control over the sub-picosecond pedestal level while remaining scalable to intensities higher than $10^{21} ~\textrm{W/cm}^2$. It relies on a simple design that positions the first plasma mirror (PM1) so that the short-delay pedestal fluence remains below $\sim 1-2 \times 10^{13} ~\textrm{W/cm}^2$ at its surface. For instance, cleaning a sub-ps contrast of $3\times 10^{-3}$ would require limiting the maximum peak intensity onto PM1 below $\sim 3-5 \times 10^{15} ~\textrm{W/cm}^2$, at the expense of reducing the transmitted energy. This energy loss can be compensated by operating the second plasma mirror (PM2) at a higher fluence ($> 2 \times 10^{16} ~\textrm{W/cm}^2$) to maximize its reflectivity. For the BELLA facility, this would require setting PM1 $\sim 30~$cm (instead of $17.5~$cm) before the focus of the $13.5~$m parabola and would result in reflectivities of $\sim 70\%$ and $\sim 88\%$ on PM1 and PM2 respectively, see Fig. \ref{fig:laser_contrast_DPM_transmission}(b), corresponding to a total DPM transmittance of $\sim 62\%$, a modest drop with respect to the $\sim 75\%$ transmittance of the present configuration at maximum power. 

\section{Fundamental limits to efficient HHG on plasma mirrors} \label{sec:PM_dynamics_HHG}
The results of section \ref{sec:exp} show that the evolution of the plasma mirror (gradient compression and surface modulation) under laser irradiation plays a central role in the loss of HHG efficiency at high intensities. We now use PIC simulations to identify the physical mechanisms driving this evolution and determine how it affects and ultimately limits the scaling of HHG with increasing laser intensity. To this end, we performed 1D and 2D simulations with WarpX v26.04 \cite{WarpXv26.04}, spanning a broad range of intensities ($10^{19}-10^{23} ~ \textrm{W/cm}^2$). As in section \ref{sec:exp}, all simulations were performed with an angle of incidence of $45^\circ$ (relying on the boosted frame technique \cite{BourdierPoF1983} for the 1D cases). We adopted a plasma gradient of $\lambda/10$, which yields optimal HHG \cite{VincentiPRL2019} at this incidence. To isolate the intrinsic evolution of the plasma mirror from pulse-envelope and finite-spot effects, we first considered a pulse with constant intensity following a short rise time of a few femtoseconds and, in 2D, a large, flat-top focal spot (see Sup. Mat. \cite{SuppMat}). The effect of the main pulse duration is investigated separately in section \ref{sec:duration}. 

\subsection {Physical mechanism of plasma mirror degradation with intensity}

To understand how plasma-mirror evolution affects HHG, we first focus on a representative 2D simulation at $I=10^{20}\,\mathrm{W/cm^2}$. This intensity is motivated by the pedestal study of section \ref{sec:exp}, where Fig.~\ref{fig:sim_constrast_effect}(c) shows that a $10^{20}\,\mathrm{W/cm^2}$ pedestal lasting $75$~fs already substantially reduces the harmonic yield of the subsequent $2\times10^{21}\,\mathrm{W/cm^2}$ main pulse. This indicates that significant plasma-mirror evolution is already induced by the pedestal on this timescale. To better follow the temporal evolution of the plasma mirror and identify the dynamics responsible for this degradation, we extend the interaction at a constant intensity of $10^{20}\,\mathrm{W/cm^2}$ up to $200$~fs in Fig.~5. As we show below, the plasma-mirror evolution proceeds through two successive stages.

\begin{figure*}[tb]
    \includegraphics[width=\textwidth]{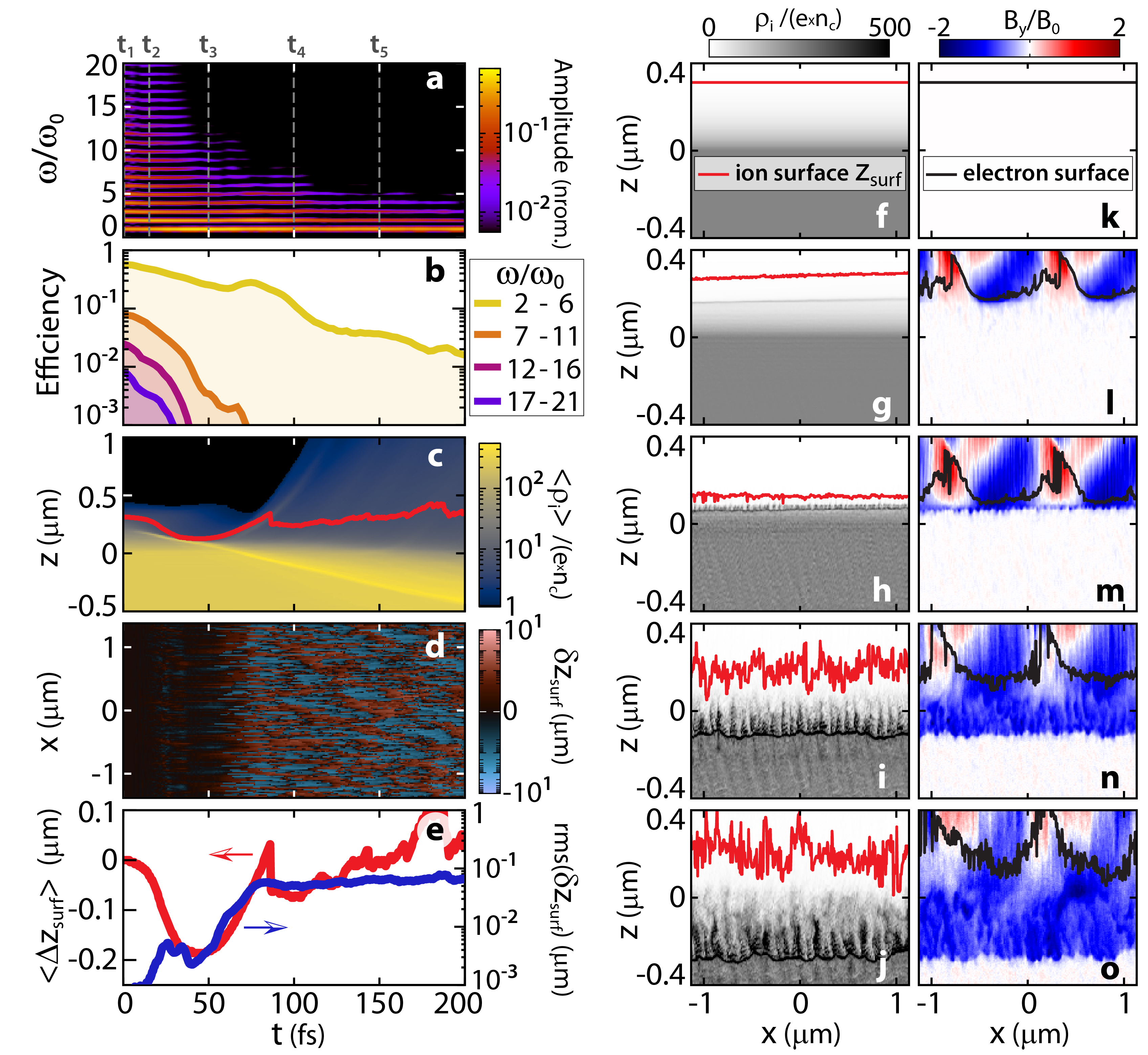}
    \caption{Evolution of the plasma mirror under irradiation by a constant-intensity pulse at $I=10^{20}\,\textrm{W/cm}^2$. (a) Time-resolved spectrogram. (b) Integrated HHG efficiency as a function of time for different harmonic-number ranges. (c) Ion density integrated over the transverse component as a function of time; the red line indicates the ion-surface position $Z_\mathrm{surf}$. (d) Evolution of the ion-surface fluctuations relatively to the average position. (e) Red (left axis): transversely averaged ion-surface position as a function of time; blue (right axis): evolution of the root-mean-square surface fluctuations. (f–j) Ion-density distributions and (k–o) the magnetic-field component transverse to the propagation plane at the time steps indicated in (a): $t_1$ -- (f,k), $t_2$ -- (g,l), etc. The red and black curves indicate the ion and electron surfaces of the plasma mirror, defined as the first position, moving from vacuum toward the target along its normal, where the electron or ion density exceeds $n_c\sqrt{1+a_0^2/2}$, where $a_0$ is the normalized peak vector potential and $n_c$ the critical plasma density (see Sup. Mat. \cite{SuppMat} for details). The harmonic signal used in (a–b) is integrated over the transverse wave-vector component. For full details of this and next figures, see the Supplementary Material \cite{SuppMat}.}
    \label{fig:overview_1e20}
\end{figure*}

The five times $t_1$--$t_5$ marked in the harmonic spectrogram of Fig .~\ref{fig:overview_1e20}(a) correspond to the snapshots of electron and ion densities shown in Figs.~\ref{fig:overview_1e20}(f--o), allowing to relate the evolution of the harmonic emission directly to that of the reflecting plasma surface.

We first consider the early-time evolution, from $t_1$ to $t_3$. The initial density gradient is progressively compressed by the laser radiation pressure, as seen from the evolution of the ion density in Figs.~\ref{fig:overview_1e20}(f--h). To quantify this evolution, we define the plasma-mirror surface $Z_{surf} $ as the first position, moving from vacuum toward the target along its normal, where the electron or ion density exceeds $n_c\sqrt{1+a_0^2/2}$, where $a_0$ is the normalized peak vector potential and $n_c$ the critical plasma density (see Sup. Mat. \cite{SuppMat} for details). This definition is analogous to that used in previous works \cite{LefebvrePRL2002, SgattoniPRE2015, GrassiPRE2017} to account for relativistic transparency effects \cite{AkhiezerJETP1956, KawPoF1970, PalaniyappanNatPhys2012, ZingaleNJP2026}. The resulting surfaces, shown by the red curves in Figs.~\ref{fig:overview_1e20}(f--j) for ions and black curves in Figs.~\ref{fig:overview_1e20}(k--o) for electrons, closely follow the region where the incident electromagnetic field is reflected, as independently evidenced by the transverse magnetic field in Figs.~\ref{fig:overview_1e20}(k--o).

During this first stage, the plasma surface is pushed into the target and the density gradient rapidly steepens. This dynamics is characteristic of hole boring, in which the laser radiation pressure acts as a piston on the overdense plasma \cite{WilksPRL1992, PukhovPRL1997, MacchiPRL2005, NaumovaPRL2009, PingPRL2012, FiuzaPRL2012, RuyerPoP2015, GrassiPRE2017, GrassiPhD2017, IwataNatComm2018, HornungNatComm2021}. The compression is accompanied by the formation, within $\sim 10$~fs, of a sharp electron--ion density peak that subsequently propagates into the target, corresponding to the formation of an electrostatic shock \cite{FiuzaPoP2013}. The temporal evolution of the integrated ion density profile in Fig.~\ref{fig:overview_1e20}(c) provides a global view of this compression and the subsequent shock propagation into the bulk.

This early evolution has a direct signature on the harmonic emission. The spectrogram in Fig.~\ref{fig:overview_1e20}(a) and the harmonic efficiencies in Fig.~\ref{fig:overview_1e20}(b) show a first rapid decrease of the HHG yield between $t_1$ and $t_3$, occurring over a few tens of femtoseconds and affecting all harmonic orders. This first drop coincides with the compression of the density gradient. Since efficient HHG requires a gradient close to the optimum value $L\simeq\lambda/10$ \cite{VincentiPRL2019}, hole boring progressively drives the interaction away from the optimal initial conditions.

A qualitatively different regime develops after $t_3$. While the plasma surface remains comparatively smooth at earlier times (Figs.~\ref{fig:overview_1e20}(f--h)), transverse modulations become clearly visible in Figs.~\ref{fig:overview_1e20}(i,j) and grow with time, ultimately evolving in a strongly rippled surface. Their development is also apparent in Fig.~\ref{fig:overview_1e20}(d), which shows the temporal evolution of the transverse surface modulations relative to the mean surface position, and is further quantified in Fig.~\ref{fig:overview_1e20}(e) by the increase of their root-mean-square amplitude after $t_3$. Concomitantly, filamentary structures develop in the transverse magnetic field (Figs.~\ref{fig:overview_1e20}(n,o)). Their growth after $t_3$ is consistent with the development of a transverse instability associated with filamentary currents, reminiscent of the electron Weibel instability reported in similar laser--plasma interactions \cite{QuinnPRL2012, RuyerPoP2015, GrassiPRE2017, SchoenwaelderNatComm2026}. Such instabilities are known to develop in the hole-boring regime \cite{FiuzaPRL2012, RuyerPoP2015, GrassiPRE2017}. At oblique incidence, the interaction additionally drives a strong current sheet along the target surface \cite{VshivkovPoP1988}, associated with the formation of a quasi-static magnetic field \cite{KatoJPlasFusRes2004, GrassiPRE2017, GrassiPhD2017}, further emphasizing the intrinsically multidimensional nature of this late-time dynamics. At still later times, following maximal compression, a blowout of the target occurs \cite{IwataNatComm2018}, with a rapidly expanding region of sub-critical plasma visible in Figs.~\ref{fig:overview_1e20}(i,j).

This second stage produces a distinct signature in the harmonic emission. After the initial rapid efficiency drop associated with gradient compression, Fig.~\ref{fig:overview_1e20}(b) shows a slower but sustained decrease in harmonic yield, while Fig.~\ref{fig:overview_1e20}(a) reveals the progressive disappearance of harmonic orders. This second regime correlates with the growth of the transverse surface modulations and magnetic filaments between $t_3$ and $t_5$. Previous studies have shown that surface corrugations can strongly degrade coherent harmonic emission \cite{DromeyNatPhys2009}, while surface modulations have also been associated with an isotropization of the harmonic emission \cite{DromeyNatPhys2006} and enhanced laser-energy absorption \cite{CerchezAppPhysLett2018}. Interestingly, we also observe the characteristic emission associated with the relativistic instability-modulated emission (RIME) regime reported by Lama\v{c} \textit{et al.} \cite{LamacPRL2023}. The underlying instability is, however, likely more complex. Additional simulations indeed show that similar transverse modulations and magnetic filaments persist for short density gradients, near-normal incidence, and densities as low as $240\,n_c$, i.e. away from the conditions favoring the Bunemann-type \cite{BunemanPR1959} instability mechanism discussed by Lama\v{c} \textit{et al.} This robustness, together with the strong magnetic filamentation observed in Figs.~\ref{fig:overview_1e20}(n,o), suggests that a Weibel-like mechanism also strongly contributes to the instability, which therefore likely results from several coupled instability processes. Fully disentangling these intertwined mechanisms is beyond the scope of the present work and will be the object of a future study. 

\begin{figure}[t]
    \includegraphics[width=\columnwidth]{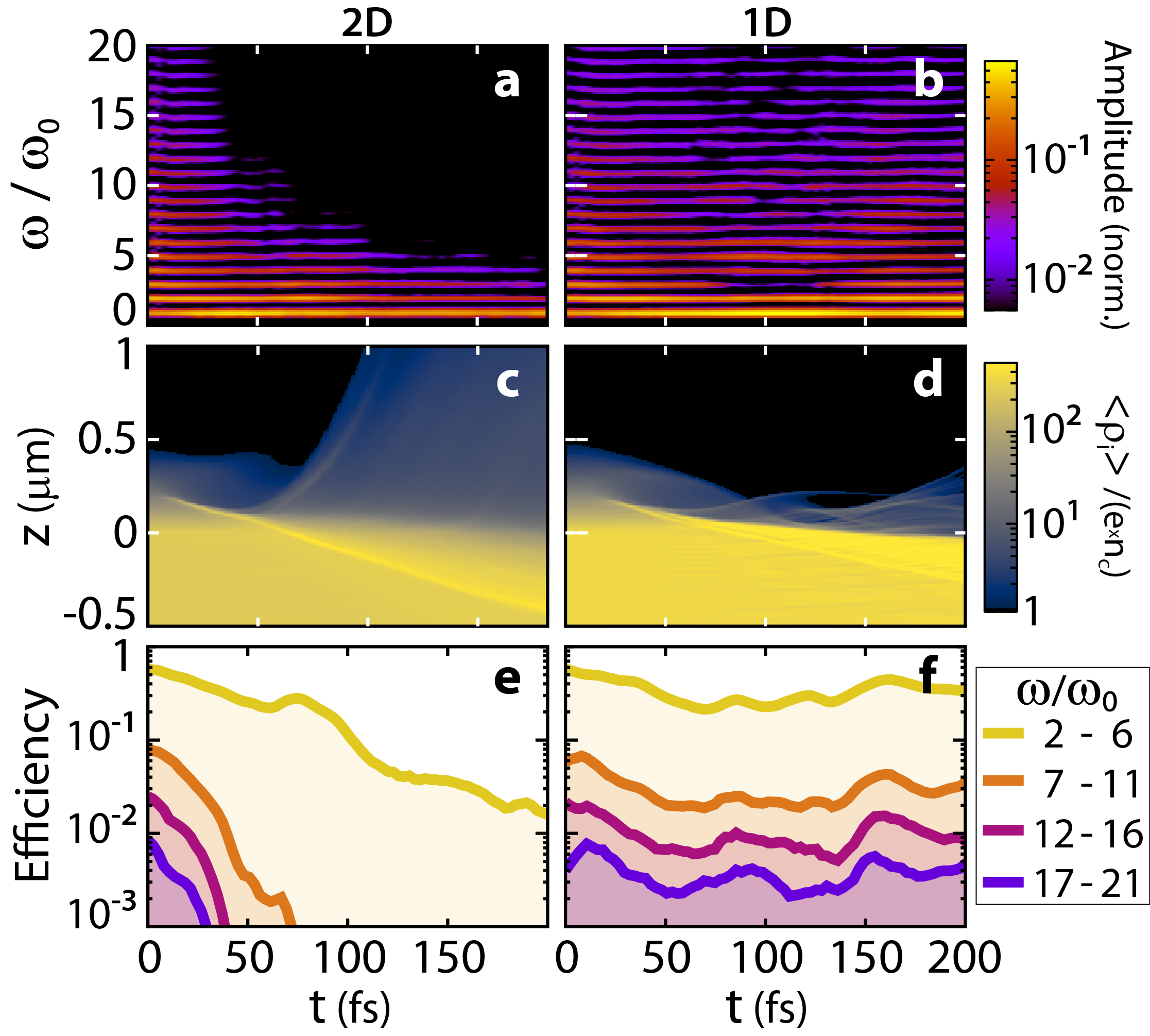}
    \caption{Comparison of 2D and 1D PIC simulations to disentangle the roles of longitudinal plasma evolution and transverse dynamics in the degradation of HHG under irradiation by a constant-intensity pulse at $I=10^{20}\,\mathrm{W/cm^2}$. (a–b) Time-resolved spectrograms. (c–d) Ion density integrated over the transverse component. (e–f) Integrated HHG efficiency for different harmonic-number ranges. The harmonic signal used in (a–b, e-f) is integrated over the transverse wave-vector component.
}
    \label{fig:2Dvs1Dvs2Dimm}
\end{figure}

Finally, to disentangle the roles of longitudinal plasma evolution and transverse dynamics in the degradation of HHG, we compare the reference 2D simulation with a 1D simulation performed with the same laser and plasma parameters (see Fig.~\ref{fig:2Dvs1Dvs2Dimm}). The 1D geometry retains the longitudinal response of the plasma to the laser radiation pressure, including hole boring and the associated compression of the density gradient, while suppressing by construction any transverse evolution of the target.

At early times (t $<$ 50 fs), the HHG efficiency exhibits a similar rapid decrease in both 1D and 2D (Fig.~\ref{fig:2Dvs1Dvs2Dimm} (a)-(b) and (g)--(h)). This behavior is consistent with the gradient-compression phase identified in section \ref{sec:PM_dynamics_HHG}: radiation pressure steepens the initially optimized density gradient away from the initial optimum. Since this process is essentially longitudinal, it is well reproduced in 1D (Fig.~\ref{fig:2Dvs1Dvs2Dimm} (e)). The observation of a similar short-time decrease in both geometries therefore supports the interpretation of this first degradation stage as predominantly resulting from the compression of the density gradient.

The subsequent evolution is markedly different. In 1D, the harmonic efficiency stabilizes after this initial decrease, whereas in 2D it continues to decrease over a longer timescale (Fig.~\ref{fig:2Dvs1Dvs2Dimm}(g)). This discrepancy coincides with the development of the surface modulations observed in the 2D simulation in Fig.~\ref{fig:overview_1e20}(d,i,j), together with the filamentary magnetic structures visible in Figs.~\ref{fig:overview_1e20}(n,o). These processes are intrinsically multidimensional, and are absent by definition in 1D. The comparison therefore provides a clear understanding of the two characteristic stages of HHG degradation: an initial, predominantly one-dimensional decrease associated with gradient compression, followed by a slower multidimensional degradation associated with the progressive corrugation of the reflecting surface. At this intensity, the two stages remain sufficiently separated in time to be individually identified. We now investigate how their characteristic timescales evolve with increasing laser intensity.

\begin{figure*}[tb]
    \includegraphics[width=\textwidth]{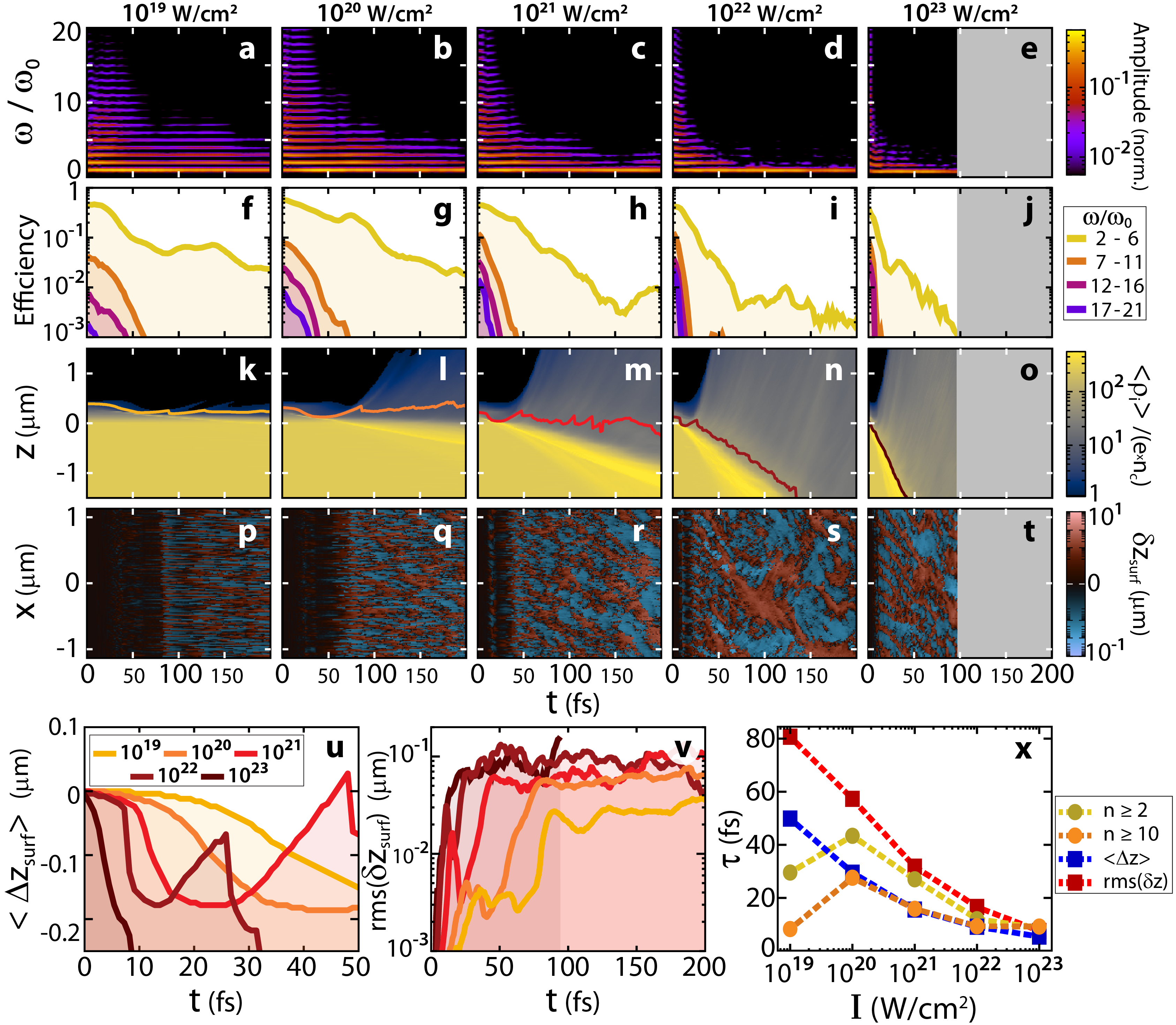}
    \caption[angle=90]{Evolution of the plasma mirror under irradiation by constant-intensity laser pulses with different peak intensities. (a–e) Time-resolved spectrograms. (f–j) Integrated HHG efficiency as a function of time for different harmonic-number ranges. (k–o) Ion density integrated over the transverse component as a function of time; the colored lines indicate the ion-surface position. (p–t) Evolution of the ion-surface fluctuations relatively to the average position. For different intensities: (u) transversely integrated ion-surface position as a function of time; (v) evolution of the root-mean-square of ion surface fluctuations. (x) Characteristic timescales for efficient generation of harmonics above 2 and 10, gradient compression $<\Delta z_\mathrm{surf}>$, and surface instability rms$(\delta(z))$. For the case of $I=10^{23}\,\mathrm{W/cm^2}$, the time dependence is limited to 100 fs due to the simulation space window limitations. The harmonic signal used in (a–j) is integrated over the transverse wave-vector component.}
    \label{fig:intensity_scan}
\end{figure*}

\subsection {Scaling of plasma mirror degradation time scale with intensity}

We now extend the analysis to the full intensity range, from $10^{19}$ to $10^{23}~\textrm{W/cm}^2$, as summarized in Fig.~\ref{fig:intensity_scan}. The harmonic spectrograms (Figs.~\ref{fig:intensity_scan}(a--e)) and efficiencies (Figs.~\ref{fig:intensity_scan}(f--j)) reveal a striking acceleration of HHG degradation with increasing laser intensity. While efficient harmonic emission persists for more than $50$~fs at $10^{19}~\textrm{W/cm}^2$, the emission window progressively shortens with increasing intensity, until HHG is suppressed within only a few femtoseconds at $10^{22}$--$10^{23}~\textrm{W/cm}^2$.

This acceleration directly reflects the increasingly rapid evolution of the plasma mirror. The evolution along the target normal (Figs.~\ref{fig:intensity_scan}(k--o)) and that of the transverse surface structure (Figs.~\ref{fig:intensity_scan}(p--t)) both accelerate strongly with intensity. At $10^{19}$--$10^{20}~\textrm{W/cm}^2$, the two stages identified in the previous section remain distinguishable: the initial compression of the density gradient precedes the development of strong transverse surface modulations. As the intensity increases, these two timescales become progressively shorter and closer to each other. Above $\sim10^{22}~\textrm{W/cm}^2$, strong transverse modulations develop while the initial plasma compression is still taking place, such that the two stages can no longer be clearly separated.

This trend is quantified in Figs.~\ref{fig:intensity_scan}(u--x), which characterize the evolution of the plasma-surface position and modulation and extract the corresponding characteristic timescales. At $10^{19}$ and $10^{20}~\textrm{W/cm}^2$, substantial plasma-mirror degradation occurs after approximately $80$ and $60$~fs, respectively. 

At $10^{22}$--$10^{23}~\textrm{W/cm}^2$, however, the situation changes qualitatively: the plasma surface becomes strongly modulated on a few-femtosecond timescale. Plasma-mirror degradation can therefore occur during the main pulse itself, even in the ideal case of a perfect 'infinite' temporal contrast. This suggests that, at sufficiently high intensity, the main-pulse duration becomes a limiting parameter for efficient HHG. We investigate this effect in the following section.

\subsection{Pulse duration threshold for efficient HHG}  \label{sec:duration}
\begin{figure*}[tb]
    \includegraphics[width=0.95\textwidth]{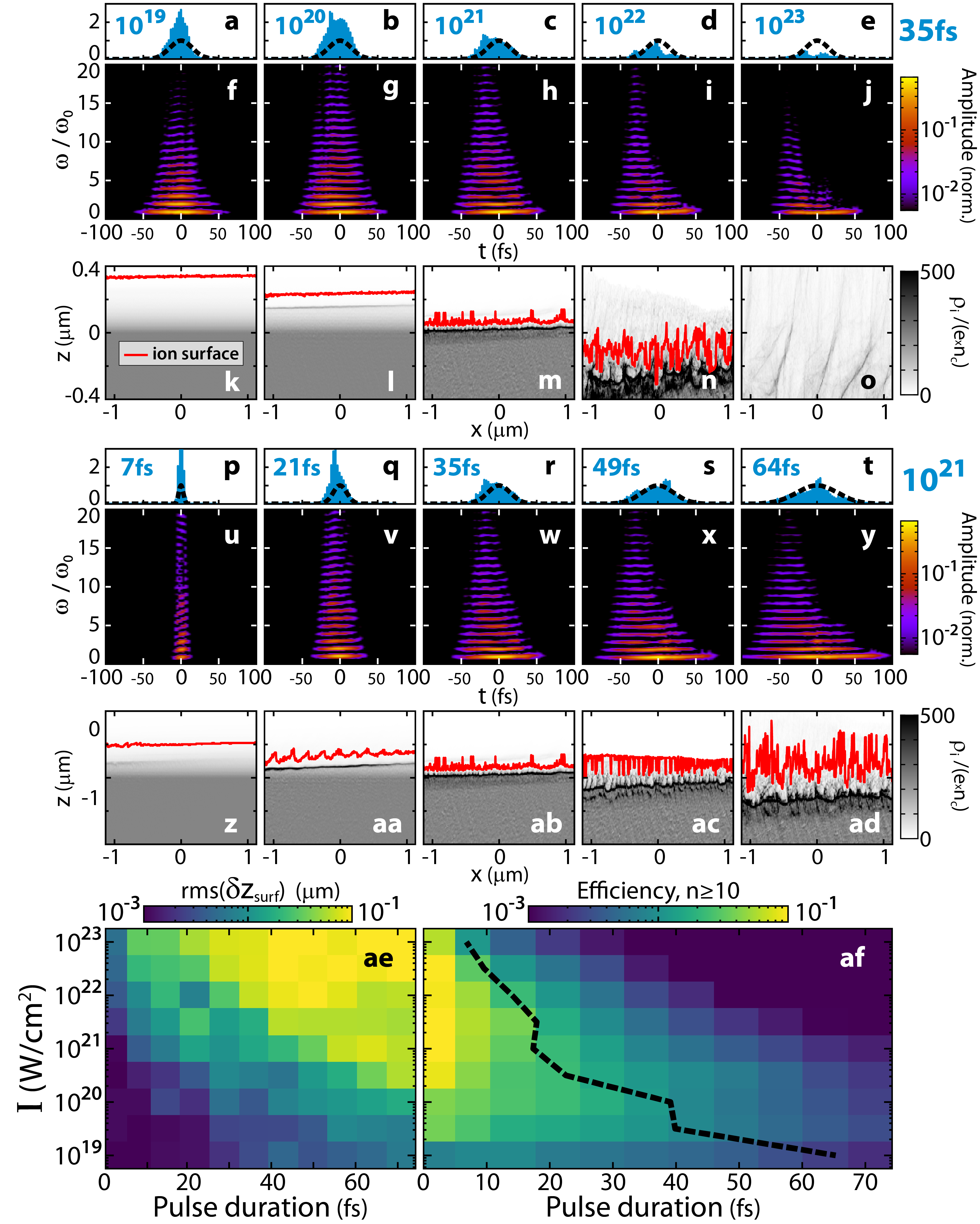}
    \caption{Effect of the pulse FWHM duration on HHG efficiency for Gaussian temporal pulse profiles. Rows 1–3: fixed pulse duration of 35 fs and different intensity. Rows 4–6: fixed intensity of $10^{21}\,\textrm{W/cm}^2$ and different pulse durations. (a–e) and (p–t) Normalized intensity of the outgoing pulse; the black dashed line indicates the envelope of the incoming pulse. (f–j) and (u–y) Time-resolved spectrograms. (k–o) and (z–ad) Ion density at the time of peak intensity on target. (ae) Root-mean-square of ion surface fluctuations as a function of laser intensity on target and pulse duratio. The firs bin corresponds to a laser pulse duration of 2fs.  (af) Integrated efficiency of harmonics above 10; the black dashed line corresponds to a drop by 25\% of the peak value for the corresponding intensity. The harmonic signal used in (a–j, p-y and af) is integrated over the transverse wave-vector component.}
    \label{fig:pulse_duration_scan}
\end{figure*}

The previous section raises an important question: even in the ideal case of perfect temporal contrast, can the plasma mirror maintain sufficient integrity over the duration of the main pulse to efficiently generate harmonics? To address this question, we performed a systematic 2D simulation campaign spanning peak intensities from $10^{19}$ to $10^{23}~\textrm{W/cm}^2$ and pulse durations from $2$ to $70$~fs, as summarized in Fig.~\ref{fig:pulse_duration_scan}.

Let us first consider a fixed pulse duration of $35$~fs (see Figs.~\ref{fig:pulse_duration_scan} (a--o)). At $10^{19}$--$10^{21}~\textrm{W/cm}^2$, the plasma surface remains comparatively smooth during the interaction and a broad harmonic spectrum is generated (Figs.~\ref{fig:pulse_duration_scan} (f,g,k,l)). As the intensity increases, modulations develop more rapidly during the pulse (Figs.~\ref{fig:pulse_duration_scan} (m--o)), accompanied by a strong reduction of the harmonic emission [Figs.~\ref{fig:pulse_duration_scan} (h--j)]. At $10^{22}$--$10^{23}~\textrm{W/cm}^2$, the plasma mirror is therefore already substantially degraded within a few 10s of fs, despite the absence of any pedestal.

Conversely, shortening the pulse strongly mitigates this degradation. Figures~\ref{fig:pulse_duration_scan} (p--ad) illustrate this effect at $10^{21}~\textrm{W/cm}^2$: for the shortest pulse, the interaction terminates before strong transverse modulations can develop, preserving an efficient harmonic emission (Figs.~\ref{fig:pulse_duration_scan} (p,u,z)). As the pulse duration increases, the surface becomes progressively more modulated ([Figs.~\ref{fig:pulse_duration_scan} (aa--ad)) and the harmonic yield correspondingly decreases (Figs.~\ref{fig:pulse_duration_scan} (v--y)).

The full parameter scan is summarized in Figs.~\ref{fig:pulse_duration_scan} (ae,af). Figure~\ref{fig:pulse_duration_scan} (ae) shows the root mean square amplitude of the plasma-surface modulation at the maximum pulse intensity, while Fig.~\ref{fig:pulse_duration_scan}(af) shows the corresponding HHG efficiency. The two maps exhibit a clear correlation: the growth of surface modulations is accompanied by a rapid loss of harmonic efficiency. Most importantly, the pulse duration for which this transition occurs (black dashed line on panel (af)) decreases strongly with increasing intensity. While pulses of several tens of fs remain compatible with efficient HHG around $10^{19}$--$10^{20}~\textrm{W/cm}^2$, increasingly shorter pulses are required above $10^{21}~\textrm{W/cm}^2$, reaching the quasi-single cycle regime at $10^{23}~\textrm{W/cm}^2$ (see left column on panel (af)).

These results reveal a second important constraint on the scaling of HHG on plasma-mirror toward extreme intensities. Temporal contrast controls the state of the plasma mirror before the arrival of the main pulse, but at sufficiently high intensity the plasma mirror evolves significantly during the main pulse itself. Efficient HHG therefore requires not only sufficient temporal contrast, but also a pulse duration shorter than the characteristic timescale for the development of surface modulations.

 While shortening the laser pulse could mitigate the development of plasma instabilities, achieving shorter pulse durations remains technically challenging at extreme laser powers. A possible alternative strategy could be to use circular polarization to reduce electron heating and potentially suppress the development of instabilities. However, circular polarization is not compatible with efficient HHG, which requires a p-polarized field. One possibility would therefore be to dynamically tailor the laser polarization, transitioning from circular to linear p-polarization near the pulse intensity peak\cite{YeungPRL2015}. This could limit plasma heating during the leading edge while preserving efficient harmonic generation around the peak intensity. Such an approach would, however, involve a trade-off, as the circularly polarized portion of the pulse would contribute little to HHG, reducing the overall laser-to-harmonic conversion efficiency. Exploring such strategies is beyond the scope of the present work and will be the subject of future investigations.

\section {Conclusion}

In conclusion, we have investigated Doppler harmonic generation from plasma mirrors at intensities substantially exceeding $10^{21}~\textrm{W/cm}^2$ and up to $10^{23}~\textrm{W/cm}^2$. As opposed to the ideal intensity scaling generally expected for HHG on plasma-mirror \cite{GordienkoPRL2005, BaevaPRE2006}, our experiments reveal a strong reduction of the harmonic yield as the laser intensity is increased. By combining the measurements with PIC simulations, we show that this collapse mainly originates from the sub-picosecond laser pedestal, which drives a significant evolution of the plasma mirror before the arrival of the main pulse. This establishes stringent sub-ps contrast requirements for extending efficient HHG to the PW regime and beyond.

Our simulations reveal the physical mechanisms underlying this degradation. Radiation-pressure-driven hole boring first compresses the initially optimized density gradient, leading to a rapid decrease of the harmonic efficiency. At later times, plasma dynamics strongly modulate the reflecting surface, resulting in a further suppression of harmonic emission. The characteristic timescales of both processes decrease rapidly with increasing laser intensity, and the distinction between them progressively disappears above $\sim10^{22}~\textrm{W/cm}^2$, where strong surface modulations develop within only few femtoseconds.

These effects introduce a fundamental constraint for HHG at extreme intensities. Even in the ideal limit of perfect temporal contrast, the plasma mirror can evolve substantially during the main pulse itself. Our pulse-duration scan shows that efficient HHG therefore requires the laser pulse to remain shorter than the characteristic timescale for the development of the surface modulations. The maximum acceptable pulse duration decreases from several tens of femtoseconds at $10^{19}$--$10^{20}~\textrm{W/cm}^2$ to just a few femtoseconds at $10^{23}~\textrm{W/cm}^2$. Temporal contrast and pulse duration thus constitute two key constraints on the operation of plasma mirrors at extreme intensities.

These results have direct implications for the use of plasma mirrors as intensity boosters for the study of the strong-field QED regime. They show that the expected scaling of harmonic properties, including conversion efficiency and spectral cutoff, breaks down at high intensity as the reflecting surface becomes unstable on increasingly short timescales. Our study provides the first comprehensive picture of the temporal-contrast and pulse-duration requirements needed to preserve the plasma mirror in the optimal regime, thereby defining the conditions for maximizing harmonic generation and the resulting intensity gain at extreme laser intensities.

\begin{acknowledgments}
We gratefully acknowledge the support of Zachary Eisentraut, Mark Kirkpatrick, Federico Mazzini, Mackinley Kath, Joe Riley, Arturo Magana, Teo Maldonado Mancuso, Chetanya Jain, Nathan Ybarrolaza, Carl Schroeder, Jeroen van Tilborg, Jens Osterhoff, Derrick McGrew, Hai-En Tsai, and Eric Esarey. \\
The work was supported by the U.S. Department of Energy’s (DOE) Office of Science (SC) Fusion Energy Sciences (FES) program: the LaserNetUS initiative at the BELLA Center, and the Office of High Energy Physics (HEP), both under Contract No. DE-AC02-05CH11231. \\
This project has received funding from the European Research Council (ERC) under the European Union's Horizon 2020 research and innovation program (grant agreements No. 101076814 and No. 101224354 - NP-QED).\\
This research used the open-source \href{https://blast-warpx.github.io}{particle-in-cell code WarpX}.
Primary WarpX contributors are with LBNL, LLNL, CEA-LIDYL, SLAC, DESY, CERN, Helion Energy, TAE Technologies, and Realta Fusion.
We acknowledge all WarpX contributors.\\
This project was provided with computing HPC and storage resources by GENCI at CINES thanks to grants 2025-A0190516880, 2025-AD010515627R1, 2024-AD010515627, and 2024-A0140513434 on the supercomputer Adastra's MI250x partition.
\\
We acknowledge the financial support of the Cross-Disciplinary Program on Instrumentation and Detection of CEA, the French Alternative Energies and Atomic Energy Commission. \\
This project has received funding from the European Union’s Horizon Europe research and innovation programme under the Marie Skłodowska-Curie Doctoral Networks grant agreement No 101169117. 
\end{acknowledgments}

\bibliography{biblio}

\end{document}